\documentclass[11pt]{article}

\usepackage{subfig}

\usepackage[tmargin=2cm, lmargin=2.5cm, bmargin=3cm,hmarginratio=1:1,headheight=65pt,headsep=1cm]{geometry} 
\usepackage[T1]{fontenc}
\usepackage[french,english]{babel}
\usepackage{color,graphicx}
\usepackage{amsmath,amssymb,amsfonts}
\usepackage{amsthm}
\usepackage{mathtools}
\usepackage{slashed}
\usepackage{sidecap} 
\usepackage{array}
\usepackage[indentafter]{titlesec}
\usepackage[utf8]{inputenc}
\usepackage{setspace}
\usepackage{perpage} 
\usepackage{fancyhdr} 
\usepackage{emptypage}
\usepackage{pifont} 

\usepackage{hyperref} 

\usepackage{color}

\newcommand{\pink}[1]{\textcolor{\pink}{#1}}

\definecolor{dblue}{rgb}{0.2,0.50,0.80}

\newcommand{\g}{\g}

\newcommand{\be}{\begin{eqnarray}}
\newcommand{\en}{\end{eqnarray}}
\newcommand{\badat}{\begin{alignedat}}
\newcommand{\eadat}{\end{alignedat}}
\newcommand{\bitm}{\begin{itemize}}
\newcommand{\eitm}{\end{itemize}}
\newcommand{\bmat}{\begin{pmatrix}}
\newcommand{\emat}{\end{pmatrix}}
\newcommand{\ba}{\begin{align}}
\newcommand{\bas}{\begin{align*}}
\newcommand{\ab}{\end{align}}
\newcommand{\bse}{\begin{subequations}}
\newcommand{\ese}{\end{subequations}}

\newcommand{\ee}{\hspace{5 mm}}

\def\L{\mathcal{L}}

\def\g{ \gamma}

\def\be{\begin{equation}}
\def\ee{\end{equation}}
\def\bea{\begin{eqnarray}}
\def\eea{\end{eqnarray}}
\def\ba{\begin{array}}
\def\ea{\end{array}}
\def\bec{\begin{center}}
\def\ec{\end{center}}
\def\ba{\begin{align}}
\def\ena{\end{align}}

\def\lab{\label}

\def\12{\frac{1}{2}}

\def\a{\alpha}

\def\g{\gamma}

\def\z{\zeta}

\def\i{\iota}

\def\l{\lambda}
\def\L{\Lambda}

\def\lp{\left(}
\def\rp{\right)}
\def\lbr{\left[}
\def\rbr{\right]}

\begin{document}

\title{\textbf{Traversable wormholes in five-dimensional Lovelock theory}}

\smallskip
\smallskip

\author{Gaston Giribet, Emilio Rub\'{\i}n de Celis, Claudio Simeone}

\maketitle

\begin{center}


{Physics Department, University of Buenos Aires and IFIBA-CONICET}\\
{{\it Ciudad Universitaria, pabell\'on 1 (1428) Buenos Aires, Argentina.}}


\end{center}


\smallskip
\smallskip
\smallskip
\smallskip
\smallskip

\begin{abstract}
In general relativity, traversable wormholes are possible provided they do not represent shortcuts in the spacetime. Einstein equations, together with the achronal averaged null energy condition, demand to take longer for an observer to go through the wormhole than through the ambient space. This forbids wormholes connecting two distant regions in the space. The situation is different when higher-curvature corrections are considered. Here, we construct a traversable wormhole solution connecting two asymptotically flat regions, otherwise disconnected. This geometry is an electro-vacuum solution to Lovelock theory of gravity coupled to an Abelian gauge field. The electric flux suffices to support the wormhole throat and to stabilize the solution. In fact, we show that, in contrast to other wormhole solutions previously found in this theory, the one constructed here turns out to be stable under scalar perturbations. We also consider wormholes in AdS. We present a protection argument showing that, while stable traversable wormholes connecting two asymptotically locally AdS$_5$ spaces do exist in the higher-curvature theory, the region of the parameter space where such solutions are admitted lies outside the causality bounds coming from AdS/CFT. 

\end{abstract}

\maketitle

\newpage

\section{Introduction}
\lab{intro}

Wormholes are one of the most fascinating solutions of gravitational field equations. Originally conceived as a hypothesis on the structure of matter in classical physics \cite{Misner:1957mt}, wormholes have served as illustrative examples on how abstruse the topology of spacetime can be \cite{Morris:1988cz}, allowing us to investigate to what extend causality, locality, and energy conditions are interrelated \cite{Fuller:1962zza}. These geometries have also provided a particularly rich source of inspiration for science fiction literature. The science fiction wormholes, however, are radically different from those seriously considered in theoretical physics; the main difference being their inviability at macroscopic scales. The existence of wormholes demands the violation of certain energy conditions, which is only possible in quantum physics.

In the last few years, the interest on wormhole geometries have been renewed. Both traversable and non-traversable wormhole geometries were recently considered in relation to quantum gravity and notably to holography \cite{Maldacena:2013xja, Chernicoff:2013iga, Gao:2016bin, Maldacena:2017axo, Maldacena:2018lmt, Gao:2018yzk, Caceres:2018ehr, Betzios:2019rds, Freivogel:2019lej, Maldacena:2018gjk}. In \cite{Maldacena:2018gjk}, an explicit example of a metastable, traversable wormhole was constructed\footnote{There have been other interesting papers recently on wormhole geometries; see for instance \cite{Ayon-Beato:2015eca, Anabalon:2018rzq, Anabalon:toappear, Horowitz:2019hgb} and references therein and thereof.}. The solution describes a pair of extremal magnetically charged black holes connected by a long throat, in such a way that it takes longer for an observer to go through the wormhole than through the ambient space, as demanded by the inalienable achronal averaged null energy condition \cite{Graham:2007va}. Einstein equations, once such energy condition on the matter fields was imposed, do not allow for wormholes with a short throat. In particular, wormholes connecting two distinct asymptotically flat regions are excluded. The situation is different when higher-curvature corrections to the gravitational action are taken into account. In that case, the higher-degree\footnote{In dimension greater than four, higher-curvature terms do not necessarily yield higher-order terms in the field equations. They can well lead to second-order field equations that are non-linear in the second derivative of the fields.} terms can effectively act as the exotic matter contribution needed for such wormholes to exist. 

Here, we will explicitly show that higher-curvature models in five dimensions do allow for stable wormholes that connect two asymptotically flat --or asymptotically Anti-de Sitter (AdS) -- regions without introducing extra exotic matter. As a working example, we will consider the Einstein-Maxwell theory supplemented with the quadratic Gauss-Bonnet higher-curvature terms; namely
\begin{equation}
S=\frac{1}{16\pi }\int d^5x\sqrt{-g}\Big( R-2\Lambda -  F_{\mu \nu} F^{\mu \nu}+\alpha (R_{\mu \nu \rho \sigma }R^{\mu \nu \rho \sigma }-4R_{\mu \nu }R^{\mu \nu }+R^2)\Big)+B\label{Uno}
\end{equation}
where $B$ stands for the boundary term that renders the variational problem well posed. This theory is the favorite model to study the effects of higher-order terms in the context of AdS$_5$/CFT$_4$ correspondence \cite{Brigante:2007nu, Hofman:2009ug}, the reason being that it is analytically solvable in many physically attractive scenarios such as AdS black holes. Besides, action (\ref{Uno}) can be motivated from string theory: it resembles the $\alpha '$-corrected low energy effective action of heterotic string theory, and it also appears in Calabi-Yau compactification of M-theory to 5 dimensions\footnote{The first higher-curvature correction of M theory in 11 dimensions is a quartic term, $R^4$ \cite{Antoniadis:1997eg, Ferrara:1996hh}. When compactifying the theory on a 6-dimensional Calabi-Yau, the effective 5-dimensional theory exhibits $R^2$ terms as those in (\ref{Uno}); see for instance Eqs. (2.6)-(2.9) in \cite{Antoniadis:1997eg}; see also Eq. (1) in \cite{Guica:2005ig}}. The specific combination of quadratic terms in (\ref{Uno}) is the only one that yields second-order field equations \cite{Lanczos:1938sf, Lovelock:1972vz}. As a consequence, the theory is free of Ostrogradsky instabilities. It also results free of ghosts around its perturbative maximally symmetric vacuum \cite{Zwiebach:1985uq}. The theory, however, exhibits some causality issues \cite{Camanho:2014apa} and other notorious features \cite{Camanho:2012da}. In \cite{Camanho:2014apa}, the causality constraints of higher-curvature models were studied, and it was shown in particular that a theory like (\ref{Uno}) has to be supplemented with massive higher-spin fields in order to be free of causality problems. Causal structure of Einstein-Gauss-Bonnet (EGB) theory has also been studied in \cite{Izumi:2014loa, Reall:2014pwa}, where different notions closely connected to causality are studied in detail, such as the relation between Killing horizons and characteristic hypersurfaces, hyperbolicity in the near horizon regions, among others. 

Theory (\ref{Uno}) is a particular case of the so-called Lovelock theory of gravity \cite{Lovelock:1971yv}, which is a natural generalization of Einstein theory to higher dimensions \cite{Lovelock:1972vz}. It is defined as the dimensional extension of a topological invariant that, in virtue of the Chern-Weil generalization of the Gauss-Bonnet theorem, computes the Euler characteristic of a 4-manifold. In such theory, we will construct traversable wormhole geometries supported by a Maxwell field without introducing exotic matter and, as a consequence, satisfying the energy conditions. Our wormhole solution connects two different asymptotically flat or AdS regions. In other words, it represents a ``short'' wormhole. The impediments that one finds when trying to construct such a solution in Einstein theory are circumvented here due to the presence of higher-curvature terms, which suffice to support the throat. This implies, in particular, that the wormhole will be microscopic, i.e. with a throat of the size $ \sqrt{\alpha }$. The geometry still represents an exact solution to (\ref{Uno}) which, in contrast to other solutions found previously \cite{Garraffo:2007fi, Richarte:2007zz, Simeone:2011zz}, is stable under S-perturbations. It is the presence of the $U(1)$ charge what stabilizes the wormhole and makes the asymptotically flat solution possible.

The fact that the wormhole throat is of order $\sqrt{\alpha }$ raises an immediate objection: At that scale, terms that are higher in the curvature would contribute and so they can not be neglected. If we think of (\ref{Uno}) as the truncation of a ultraviolet complete theory, the effective theory would breakdown at that scale. In addition, we will see that the geometric construction of the higher curvature wormholes requires to consider in certain internal patches of the spacetime solutions to (\ref{Uno}) that are non-perturbative, in the sense of having $S_{\text{on-shell}}\sim 1/\alpha$ --these solutions are somehow analogous to the self-accelerated solutions of higher-derivative models --. However, one can in principle answer to these objections in the following way: Despite higher-curvature terms can not be neglected at length scales $\sqrt{\alpha }$, one has the expectation that the qualitative features introduced by the terms quadratic in the curvature will not be drastically affected by the introduction of, say, quartic or higher terms. More precisely, one can imagine a window in the parameter space such that the wormhole throat, although microscopic, is still larger than the length at which quartic terms become relevant. We will adopt here a pragmatic point of view: As in other works in which higher-curvature actions like (\ref{Uno}) are considered --notably in the context of AdS$_5$/CFT$_4$ --, and even when such actions seem to violate the logic of effective field theory and require fine tuning, we will take into account that holography does not exclude such models and so they might in principle be considered \cite{Belin:2019mnx}. 

We should probably add that wormhole solutions in 5-dimensional EGB theory (\ref{Uno}) have already been discussed in the literature \cite{Dotti:2006cp, Dotti:2007az, Maeda:2008nz, Correa:2008nq, Kokubu:2015spa}. Our solutions, however, differ from those considered before in many aspects. For example, the main differences between the wormhole solutions considered in \cite{Dotti:2006cp, Dotti:2007az, Correa:2008nq} and ours are the following: First, the solutions considered therein only exist at the so-called Chern-Simons point $\Lambda\alpha=-3/4$ \cite{Zanelli:2005sa}, while in our solutions exist for a continuous range of the parameter $\alpha $ after the cosmological constant has been fixed. Second, the base manifold in the solution of \cite{Dotti:2006cp, Dotti:2007az, Correa:2008nq} is different from those of our examples: while we consider 3-dimensional constant curvature manifolds for the constant-time, constant-radius foliation of the geometry, the base manifold of the solution considered in \cite{Dotti:2006cp, Dotti:2007az, Correa:2008nq} is the product of a 2-dimensional locally hyperbolic space $H_2/\Gamma$ and a circle $S^1$. A third difference is that our solutions are constructed by gluing together different patches, all of them corresponding to different values of $\xi$ in the spherically symmetric solution (\ref{BD}) below. In contrast, the solution of \cite{Dotti:2006cp, Dotti:2007az, Correa:2008nq} is given by a smooth $C^{\infty }$ single manifold --at the Chern-Simons point, the different effective cosmological constants of the higher-curvature Lovelock theory coincide--. Our solutions also differ from the class of solutions studied in \cite{Maeda:2008nz}. There, the authors find some conditions and no-go propositions for the existence of wormhole type solutions in EGB theory coupled to matter. Those results are certainly consistent with our finding: For instance, in the proposition 2 of \cite{Maeda:2008nz}, the point is made that the null energy condition has to be violated if the wormholes throat embedded in the so-called general relativity (GR) branch of EGB theory (which corresponds to the choice $\xi =1$ in (\ref{BD}) below). Our solution does obey the null energy condition on the throat precisely because its throat is embedded in the opposite branch (the branch $\xi =-1$ of (\ref{BD}) below). Related to this, it is important to mention that the solutions we consider here also differ from those considered in reference \cite{Garraffo:2007fi}, where non-GR branches were also considered: In contrast to those in \cite{Garraffo:2007fi}, the solutions we construct here have attached an asymptotic region with arbitrary small effective cosmological constant. In addition, the presence of the electric flux renders our solutions stable under scalar perturbations, unlike those in \cite{Garraffo:2007fi}; this also makes our solutions different from those analyzed, for instance, in \cite{Kokubu:2015spa}.


This paper is organized as follows: In section 2, we will consider the spherically symmetric, static solutions to the higher-curvature theory coupled to a Maxwell field. These solutions will be the building blocks of our geometry, while the blinder agent will be the junction conditions derived from the boundary term $B$ in (\ref{Uno}). These conditions are the generalization of the Israel junction conditions of general relativity. We will discuss this also in section 2. In section 3, we will construct the wormhole by assembling different patches of the spacetime. We will do this without resorting to exotic matter. The electric field and the higher-curvature terms will suffice to support the wormhole throat. In section 4, we will study the stability conditions for our wormhole solution. We will show that, in certain regions of the parameter space, the solution turns out to be stable under scalar perturbations. In section 5, we will generalize the solution to the case of asymptotically AdS$_5$ wormholes, and we will make some comments in relation to AdS/CFT correspondence.

\section{Higher curvature gravity}
\lab{hcg}

Let us begin by considering the black hole solutions of the theory defined by (\ref{Uno}), namely \cite{Boulware:1985wk, Wiltshire:1988uq}
\be \lab{metric}
ds^2 = g_{\mu \nu }dx^{\mu }dx^{\nu } = - f(r) dt^2 + f^{-1}(r)dr^2 + r^2 d\Omega^2 
\ee
where $d\Omega^2$ is the metric on a 3-space of constant curvature $k=0,\pm 1$, and where the function $f$ is given by
\be
f(r) \lab{BD}
= k + \frac{r^2}{4 \a}  \lp 1 + \xi  \sqrt{1 +  \frac{16 \a M}{r^4} -  \frac{8 \a Q^2}{3 r^6} + \frac{4 \a \Lambda}{3}} \rp
\ee
where $\xi =\pm 1$. Here, $t\in \mathbb{R}$, $r\in \mathbb{R}_{\geq 0}$. The solution with $\xi = -1$, $k=1$ and $\Lambda=0$, in the large $r $ limit, tends to the 5-dimensional Reissner-Nordstr\"om solution of Einstein-Maxwell theory. The non-zero components of the spherically symmetric electromagnetic field are $F^{t r} = {Q}/{r^3}$, with $Q$ being the electric charge. We consider\footnote{This is the sign compatible  with string theory.} $\a>0$ and, mainly, asymptotically flat spacetimes which correspond to $\xi=-1$ solutions and vanishing cosmological constant. These are $k=1$ spherical geometries with branch singularity at $r=r_S$, shielded by two horizons for charges $|Q|<Q_c=\sqrt{3}|M-\a|$, with $M$ the ADM mass. If $|Q|=Q_c$ there is one horizon only and for $|Q|>Q_c$ there is a naked singularity at the radius where the radicant in (\ref{BD}) vanishes, i.e. at the surface $r=r_S$, where $r_S$ is the largest real solution to the equation $r_S^6(1+4\alpha\Lambda/3)+r_S^2 16\alpha M-8\alpha Q^2/3=0$. The outer horizon for $|Q|<Q_c$ is at $r=r_H$, where
\be
r_H^2 = \frac{Q_c}{\sqrt{3}}+ \sqrt{\frac{Q_c^2-Q^2}{3}} \,.
\ee

As said in the introduction, our wormhole will be constructed by joining together different patches of the geometry (\ref{metric}). As junctures, we consider timelike shells which separates two bulk geometries with some $f_{-}(r)$ and $f_{+}(r)$ as in (\ref{BD}). These shells are described by radial coordinate $r=\rho(\tau)$ and time coordinates $t_{\mp} = T_{\mp}(\tau)$, where the parameter $\tau$ is the proper time on the shell \cite{Poisson:1995sv}.
The induced metric $dh^2 = -d\tau^2 + \rho^2(\tau)  \, d\Omega^2$ is the same from both sides and then $f_{\mp} \dot{T}_{\mp}^2 - f_{\mp}^{-1} \dot{\rho}^2 = 1$, at the shell.
%
The induced stress tensor $S_{ij}$ at the hypersurface is given by
\be \lab{stress_tensor}
- 8\pi  S_{ij} = \Big( K_{ij}-K h_{ij} + 2\alpha ( 3J_{ij}-Jh_{ij}+2P_{iklj}K^{kl}) \Big)\Big|^{+}_{-} ,
\ee
where Latin indices label the coordinates on the joining surface of induced metric $h_{ij}$, $\mathcal{F}|^{+}_{-}  \equiv \mathcal{F}_+ - \eta \mathcal{F}_-$, with $\eta=\pm 1$ depending on the relative orientation in both sides of the shell, $K_{ij}$ is the extrinsic curvature tensor over the hypersurface at each side, and the divergence-free part of the Riemann tensor $P_{ijkl}$ and the tensor $J_{ij}$ are defined as
\be
P_{ijkl}=R_{ijkl} + R_{jk}h_{li} - R_{jl}h_{ki} - R_{ik}h_{lj} + R_{il}h_{kj} + \frac{1}{2}\, R\, (h_{ik}h_{lj}-h_{il}h_{kj}),
\ee
\be
J_{ij}=\frac{1}{3} \lp 2KK_{ik}K^{k}_{j}+K_{kl}K^{kl}K_{ij}-2K_{ik}K^{kl}K_{lj}-K^{2}K_{ij} \rp \,.
\ee
Junction conditions (\ref{stress_tensor}), which follows from the boundary term $B$ in (\ref{Uno}), are the generalization of the Israel conditions to the case in which higher-curvature terms are included \cite{Davis:2002gn}.

\section{Wormhole solution}
\lab{whs}

We will construct electro-vacuum wormhole solutions to (\ref{Uno}) with two asymptotic regions. The spacetime is symmetric across the throat and outside the mouth of the wormhole the solution corresponds to a charged, static black hole geometry. We mainly focus on the case of asymptotically flat configurations, but we keep track of a non vanishing cosmological constant to later extend it to the case of wormholes in AdS$_5$ and make some comments in relation to AdS/CFT.

Our solution is constructed with four bulk pieces, four distinct patches joined by three 4-surfaces. 
Each piece is a region obtained from 
metric (\ref{metric}) by removing part of the geometry on one side of some codimension-one hypersurface defined at fixed radial coordinate. The four bulk regions are pasted to construct the geodesically complete manifold $\mathcal{M}$ describing our traversable wormhole. 
We denote the four regions as left exterior $\mathcal{M}_L^{e} $, left interior $\mathcal{M}_L^{i}$, right interior $\mathcal{M}_R^{i} $, and right exterior $\mathcal{M}_R^{e}$, with their respective coordinates ${x^{\alpha}_{L,R}}$ for the {\textit{Left}} (L) and {\textit{Right}} (R) bulks. Explicitly, they are
\be 
\mathcal{M}_L^{e} = \{ x^\alpha_L \, / \, r_L \ge b \}, \qquad 
\mathcal{M}_L^{i}=\{ x^\alpha_L \, / \,  b \ge r_L \ge a \} ,\nonumber
\ee
\be 
\mathcal{M}_R^{i}=\{x^\alpha_R \, / \,  a \le r_R \le b \}, \qquad 
\mathcal{M}_R^{e}=\{x^\alpha_R \, / \, b \le r_R \}\,,\nonumber
\ee
and the complete manifold is $\mathcal{M}=\mathcal{M}_L^{e}  \cup \mathcal{M}_L^{i}  \cup \mathcal{M}_R^{i}  \cup \mathcal{M}_R^{e} $.
Each external region $\mathcal{M}_{L,R}^e$ is joined to its corresponding inner region $\mathcal{M}_{L,R}^i$ at the hypersurface of a bubble $\Sigma_{L,R}^b$. In other words, $\Sigma_{L,R}^b=\mathcal{M}_{L,R}^i \cap \mathcal{M}_{L,R}^e$. The {\textit{Left}} bubble is placed at $r_L=b$ and the {\textit{Right}} side bubble, symmetrically, at $r_R=b$. 
Inner regions are also glued to each other at the throat located at $r_L = a = r_R$, with $a < b$; i.e. $\Sigma^a=\mathcal{M}_{L}^i \cap \mathcal{M}_{R}^i$. This is depicted in figure (\ref{wormhole}).
The junction hypersurfaces are described as $\Sigma_L^b = \partial \mathcal{M}_L^{e}  = \partial \mathcal{M}_L^{i}|_b$ 
and $\Sigma_R^b  = \partial \mathcal{M}_R^{e}  = \partial \mathcal{M}_R^{i}|_b$ 
which define bubbles, and 
$\Sigma^a = \partial \mathcal{M}_L^{i}|_a  = \partial \mathcal{M}_R^{i}|_a$ 
which corresponds to the wormhole throat. 

\begin{figure} [h!] 
\centering
{\includegraphics[width=13.0cm]{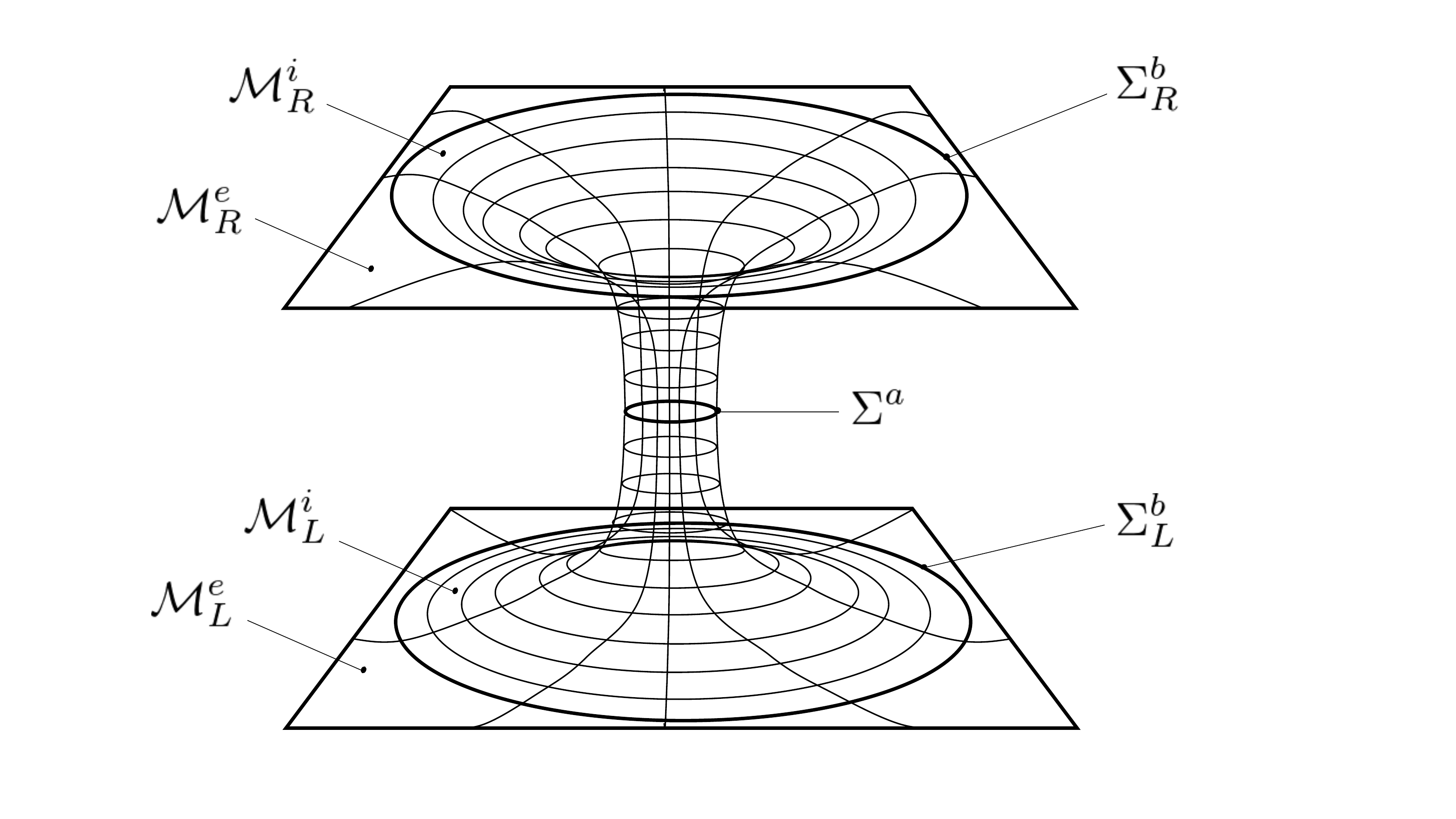}}
\caption{Scheme of the wormhole geometry.} 
\lab{wormhole}
\end{figure}
The metric function is $f_i(r_{L,R})$ for interior geometries and $f_e(r_{L,R})$ for the exteriors, with mass parameters $M_{i} = M_{Li} = M_{Ri}$ and $M =  M_{Le} = M_{Re}$, respectively.
The exterior metrics must belong to the general relativity branch $\xi_{e} = -1$ and have vanishing cosmological constant in order to have flat asymptotics and, consequently, we set $\L=0$.
The hypersurface at the throat determines that the inner regions must both be of the $\xi_{i} = +1$ branch to construct the symmetric wormhole \footnote{A static throat with null or positive energy density (and $\a>0$) is incompatible with bulk regions of the $\xi=-1$ branch at both sides of the hypersurface \cite{Garraffo:2007fi}.}.
The configuration implies a throat whose radial coordinate is greater than the branch singularity of the interior metrics, $a>r_{S_i}$, while bubbles radii must be greater than the would be horizons of the external black hole metrics or their singularity surfaces; $b > \mbox{Max}\{ r_{S_e}, \,r_H \} $.

We are looking for solutions without matter sources, so that the construction requires zero energy density and pressure at the bubbles and throat, implying vanishing induced stress tensor at the shells. 
A Gauss law integration of Maxwell equations across the non-charged 
surfaces of the shells, $\oint F_{\mu \nu} \, dX^{\mu \nu} = 0$ with $dX^{\mu \nu} = r^{ [\mu} t^{\nu]} d\Omega$, gives the continuity of the electromagnetic field in the temporal $t^{\mu}$ and radial $r^{\nu}$ directions, as
\be
\lp \eta F_{\mu \nu} t^{\mu} r^{\nu} \rp\Big|^{+}_{-} 
=0
\,,\ee
where $+$ and $-$ indicates the bulks at each side of the shell, with the orientation factor $\eta$ of each bulk region been $\eta=+1$ for radial coordinates pointing from \textit{Left} to \textit{Right} in the wormhole manifold, or $\eta=-1$ if the radial coordinates points in the opposite direction. 
The latter condition can be given in an orthonormal frame as 
$(\eta \, F_{\hat{t} \hat{r}} )_{+} = (\eta \, F_{\hat{t} \hat{r}} )_{-} $ 
or, using the metric function $f=-g_{tt} =g_{rr}^{-1}$, directly as $\lp \eta \, F_{t r}  \rp_{+} = \lp \eta \, F_{t r}  \rp_{-} $.
From the continuity of the electromagnetic field across the shells, it is seen that the charges are $Q=Q_R=-Q_L$.

Junction conditions for a generic timelike hypersurface separating two bulk regions with metric functions $f_-$ and $f_+$, 
located at radial coordinate $r= \tilde{\rho} \equiv \rho(\tau)$, establish a diagonal stress tensor (\ref{stress_tensor}) at the shell with components given explicitly by:
\begin{flalign} \lab{energy_dens}
S_{\tau}^{\tau} =  \eta \, \frac{\a}{2 \pi \, \tilde{\rho}^3} \, \sqrt{f(\tilde{\rho}) + \dot{\tilde{\rho}}^2} 
\lbr  3 \lp k + \, \frac{\tilde{\rho}^2}{4 \a} \rp - f(\tilde{\rho}) + 2 \, \dot{\tilde{\rho}}^2 \rbr  \Bigg|^{+}_{-} \,,&&
\end{flalign}
\begin{flalign} \lab{pressure}
S_{\theta}^{\theta} = S_{\phi}^{\phi} = S_{\chi}^{\chi} =  & \;
\eta \, \frac{1 }{4 \pi \, \tilde{\rho} \, \sqrt{f(\tilde{\rho}) + \dot{\tilde{\rho}}^2}}  \, \times & \\
& \lbr \frac{\a}{\tilde{\rho}} \lp k + \frac{\tilde{\rho}^2}{4 \a} - f(\tilde{\rho})  \rp  f'(\tilde{\rho}) + f(\tilde{\rho}) + \dot{\tilde{\rho}}^2 + 
 \frac{ 2 \a \ddot{\tilde{\rho}}}{\tilde{\rho}} \lp f(\tilde{\rho}) + 2 \dot{\tilde{\rho}}^2 + k + \frac{\tilde{\rho}^2}{4\a} \rp  \rbr 
 \Bigg|^{+}_{-} \, , & \nonumber
\end{flalign}
where the dots stand for derivatives with respect to $\tau $. For a static timelike shell at $r = \rho$, such that $\dot{\tilde{\rho}}=0$ and $\ddot{\tilde{\rho}}=0$, we obtain
\begin{flalign} \lab{energ_static}
S_{\tau}^{\tau} = & \frac{\a}{2 \pi \rho^3} \lp \eta_{+} \sqrt{f_+(\rho)}  -  \eta_{-} \sqrt{f_-(\rho)} \rp
\lbr 3 \lp k +  \frac{\rho^2}{4\a} \rp -  f_+(\rho) - f_-(\rho) - \eta_{+} \eta_{-} \sqrt{f_+(\rho) f_-(\rho)} \rbr ,&
\end{flalign}
\begin{flalign} \lab{pres_static}
S_{\theta}^{\theta} = S_{\phi}^{\phi} = S_{\chi}^{\chi} 
=  & \lp \frac{\eta_{+}}{\sqrt{f_+(\rho)}} - \frac{\eta_{-}}{\sqrt{f_-(\rho)}}\rp
 \lbr  k  - \z(\rho)  -  \sqrt{f_+(\rho) f_-(\rho)} \, \eta_{+}\eta_{-} \rbr \,, &
\end{flalign}
where we used the definition
\begin{flalign} \lab{zeta}
&\z(\rho) \equiv  \;  - \frac{2\a}{\rho} \lp  k  + \frac{\rho^2}{4 \a} -f(\rho) \rp  f'(\rho) + k -f(\rho) 
=  \; \frac{Q^2}{3 \rho^4} + \frac{\rho^2 \Lambda}{3} \,,&
\end{flalign}
and in the last equality of (\ref{pres_static}) it is assumed that $Q^2$ and $\Lambda$ are the same at both sides of the hypersurface. We recall that, despite we are mainly interested in $\L=0$ and $k=1$, we keep track of cosmological constant solutions and curvature parameter $k$ in some general expressions to be able to compare cases, and for comments on section \ref{AdS}.

We first consider vanishing stress tensor components in the static case to apply them to the construction of the throat. The inner regions of our wormhole have metric functions $f_{i}(a)$ 
at the position of the throat, at $r_L=r_R=a$, and the orientation factor for each bulk region at the throat is given by $\eta_{Li} = -1$ and $\eta_{Ri} = 1$. Putting all together in 
(\ref{energ_static}) and (\ref{pres_static}) the following two equations are obtained:
\be \lab{energ_a}
f_{i}(a)
= 
3 \lp k +  \frac{a^2}{4\a} \rp \,,
\ee 
\be \lab{pres_a}
f_{i}(a) = \z(a) - k \,.
\ee
On the other hand, at the \textit{Left} and \textit{Right} side bubbles placed at $r_L=b$ and $r_R=b$, the exterior regions have metric functions $f_{e}(b)$, 
 while the \textit{Left} and \textit{Right} interior regions have $f_{i}(b)$.
The orientation factor for the bulk regions at each bubble is given by:
$\eta_{Li} = -1$ and $\eta_{Le} = -1$ for the \textit{Left} side bubble, and
$\eta_{Re} = 1$ and $\eta_{Ri} = 1$ for the \textit{Right} side bubble.
By demanding 
vanishing stress tensor, the following two equations are obtained
\be \lab{energ_b}
\sqrt{f_{e}(b)} \sqrt{ f_{i}(b)} 
 = 3 \lp k +  \frac{b^2}{4\a} \rp -  f_{e}(b) - f_{i}(b) \,,
\ee
\be \lab{pres_b}
\sqrt{f_{e}(b)} \sqrt{ f_{i}(b)} =  k - \z(b) \,.
\ee
The latter four equations determine the possible configurations for the wormhole spacetime. A priori, from the simultaneous requirements in (\ref{pres_a}) and (\ref{pres_b}), we see that there are no solutions with vanishing charge and vanishing cosmological constant. Considering non vanishing charge only; $k=-1$ and $k=0$ curvature are not admissible either. The remaining possibility with $\L=0$ and $Q\neq0$, are the spherical ($k=1$) wormholes which are shown to exist and are studied below. The inclusion of $\Lambda \neq 0$ generates a variety of possibilities commented in the last section.

Now, let us study the space of solutions. To analyze the wormhole construction we use the following definitions for the configuration parameters
\be
x \equiv \frac{a^2}{4\a} \;, \qquad 
y \equiv \frac{b^2}{4\a} \;,
\ee
\be 
m_i \equiv \frac{M_{i}}{\a} \;, \qquad 
m_{} \equiv \frac{M_{}}{\a}  \;,
\ee
\be
q^2 \equiv \frac{1}{3} \lp \frac{Q}{4\a} \rp^2  \;, \qquad
\l \equiv \frac{4 \a \L}{3} \;.
\ee
Squaring conveniently equation (\ref{energ_a}), and combining it with (\ref{pres_a}), the static equations for a general symmetric vacuum throat are:
\be \lab{P_A}
4 k x^2  + (4 k ^2 - m_i) x + 3 q^2 = 0 \,,
\ee
\be \lab{P_B}
 3 (3-\l) x^2 + 16 k x  +  (4 k^2 - m_i) = 0\,,
\ee
with the condition $x+k>0$, not to lose the information of the sign in the original unsquared equation.
Additionally, the requirement $x>x_s$, where $x_s=(r_{S_i})^2/(4\a)$ is the branch singularity corresponding to the interior radial coordinate $r_{S_i}$ must be considered.
The latter equations establish the relation between $x$, $m_i$, $q$ and $\l$ compatible with a generic symmetric vacuum throat. 

It is worth pointing out that the existence of the wormhole solutions demands the presence of the higher-curvature terms. This is why the GR limit $\alpha \to 0$ in the equations above does not yield any configuration of this class. Both finite values of the higher-curvature coupling $\alpha $ and the electric charge $Q$ are necessary for having solutions like the ones we presented here. Nevertheless, the role played by the higher-curvature terms and by the charge are different: while the former are needed and are sufficient to have wormhole solutions in 5-dimensional EGB theory \cite{Garraffo:2008hu}, the latter is what, as we will see below, suffices to render the solution stable under radial perturbations, cf. \cite{Garraffo:2007fi}.

On the other hand, the vacuum bubbles at radial coordinates $b$ are constructed from the junction of an interior metric with
$f_{i}(b) = k + y ( 1 + I_y)$ 
and an exterior 
$f_{e}(b) = k + y ( 1 - E_y)$, where
$I_y \equiv  (1+ \frac{m_i}{y^{2}} - \frac{2 q^2}{y^{3}} + \l)^{1/2}$ and
$E_y \equiv  (1+ \frac{m_{}}{y^{2}} - \frac{2 q^2}{y^{3}} + \l)^{1/2}$, respectively. 
From equations (\ref{energ_b}) and (\ref{pres_b}) of the shell we read the general conditions $k+y>0$ and $k>\z_y$, with $\z_y \equiv \z(b) =  (q/y)^2 + y \l $. Additionally, after some manipulations, we have the inequalities $I_y>E_y>I_y/2$. Considering these conditions, the combined vacuum bubble equations can be expressed as
\be \lab{static_pot_y}
 y \, I_y \, E_y =  \lbr 3 (k + y)  - y (I_y - E_y) \rbr (I_y - E_y)  \,,
\ee
\be \lab{zeda_y}
\z_y + y= y (I_y - E_y) \,.
\ee
We stress that $y > y_h = r_H^2/(4\a)$ if the charge is lower than the critical value for the exterior metric, or if not, then $y > y_s = (r_{S_{e}})^2/(4\a) $ to avoid branch singularity.
The latter equations and conditions give the relations between $y$, $m_i$, $m$, $q$ and $\l$ compatible with the static vacuum bubble.
Note that the positivity of each side in (\ref{zeda_y}) is fixed by the aforementioned conditions and, therefore, $M<M_i$ for the construction of the bubbles.  


We are mainly interested in the spherical $k=1$ solutions with $\l=0$, the static throat configurations in this case 
are described by 
\be \lab{q_x}
q^2 = 4 x^2 + 3 x^3 \,,
\ee 
\be \lab{mi_x}
m_i=4+16 x+9 x^2 \,.
\ee
Evaluating with $\l=0$ and $k=1$, the condition $k > \z_y = (q/y)^2$ becomes an inequality for charge and bubble radius which reads
\be
|q|<y \,,
\ee
and equations (\ref{static_pot_y}) and (\ref{zeda_y}) give $m_i$ and $m$ in terms of $q$ and $y$
\be \lab{mi_yq}
m_i 
=  \frac{1}{2} 
\lp 6 y + 3 y^2 +  (6 + 8y) \frac{q^2}{y^2}  - \frac{q^4}{y^4} + \lbr 3 \lp 4 + 3 y - \frac{q^2}{y^2} \rp \lp y + \frac{q^2}{y^2} \rp^3 \rbr^{1/2}  \rp \,,
\ee
\be \lab{m_yq}
m
=  \frac{1}{2} 
\lp 6 y + 3 y^2 +  (6 + 8y) \frac{q^2}{y^2}  - \frac{q^4}{y^4} - \lbr 3 \lp 4 + 3 y - \frac{q^2}{y^2} \rp \lp y + \frac{q^2}{y^2} \rp^3 \rbr^{1/2}  \rp \,.
\ee
Combining the latter functions with those obtained for $m_i$ and $q^2$ in terms of $x$, from the shell in the throat (\ref{q_x})-(\ref{mi_x}), we establish our vacuum wormhole solutions in parameter space. The compatible configurations are shown in the figures presented in the next section, with the corresponding stability analysis.

Small wormholes with $\l=0$ can be studied  assuming the existence of configurations with $x$ {\small $\lesssim$} $y \ll1$. From (\ref{q_x}) and (\ref{mi_x}), the interior mass and charge in this approximation are $m_i \simeq 4 + 16 x $ and $q^2 \simeq 4 x^2$. 
Under this consideration $\z_y = (q/y)^2 \simeq (2x/y)^2$.
Besides, considering the inequality $E_y>I_y/2$, we have 
\be
m  >  1 + 4 x + 6x^2/y  + \frac{3}{2} \lp 3x^2/2  - y^2/2 + 3x^3/y    \rp \,, 
\ee
for the exterior mass. Using the latter, and Eqs. (\ref{mi_yq}) and (\ref{m_yq}) to express the sum $m_i + m $ to first order in the small parameters we have, 
\be \lab{omega}
 - {\z_y}^2 + 6 \, \z_y + y \, \theta  \, \gtrsim \, 5 
\ee
where $\theta =  \frac{13}{2} \z_y - 10 \, {\z_y}^{1/2} + 6 \sim \mathcal{O}(1)$, and positive, for $0< \z_y <1$.
Defining $\epsilon$ as a positive quantity of the same order as $x$, $y$ and $q$ in the small wormhole, the inequality (\ref{omega}) establishes that $1>
{\z_y}^{1/2} > 1 -\epsilon  
$, which is consistent with the assumptions and conditions. Finally, under this approximation,
\be
2 x \simeq y \, {\z_y}^{1/2} =  y  - \mathcal{O}(\epsilon^2) 
\ee
and $q^2 = y^2 - \mathcal{O}(\epsilon^3)$. Using the latter to evaluate (\ref{m_yq}) we obtain
\be
m = 1 + 4 |q| - \mathcal{O}(\epsilon^2) \,.
\ee
The charge is $|q| = \frac{m -1}{4} + \mathcal{O}(\epsilon^2)$ and, in terms of the original parameters of the metric, a small wormhole solution is compatible with a charge greater but approximately equal to the critical value, $|Q|/\alpha \simeq \sqrt{3}  (M/\alpha - 1)$, of the external black hole geometry. These solutions are shown in the small parameter regions of the figures in next section, i.e., small dimensionless radii of the shells, $a/\sqrt{\alpha}$ and $b/\sqrt{\alpha}$, in figures \ref{b_vs_a} and \ref{M_vs_a}, and small dimensionless charge $|Q|/\alpha$ in figure \ref{Q_vs_M}.

\section{Stability analysis}
\lab{stability}

The dynamics of throat and bubbles is determined from the junction conditions by analyzing the radii $a(\tau)$ and $b(\tau)$ as dynamical variables, introducing small perturbations around the equilibrium. We will consider the method originally introduced in \cite{Brady:1991np} to study the stability of thin shells in GR. This amounts to find an effective potential for the variables $a(\tau)$ and $b(\tau)$ and then study the conditions for its convexity. This method of studying the thin shells ``bounded excursion'' has been considered extensively in the GR literature --see for instance subsection 4.1 of \cite{Visser:2003ge}-- and in the context of wormholes it has been originally considered in \cite{Poisson:1995sv}.
Here, we will first consider the stability of the throat alone, and then add the analysis for the bubble to determine the stability of the complete configuration. 

The dynamics of the throat at radius $\tilde{a} \equiv a(\tau)$ is described by $\dot{\tilde{a}}^2+ V(\tilde{a}) = 0$, with the effective potential
\be
V(\tilde{a}) = k + \tilde{x} -  \tilde{x} \, \frac{\tilde{I_{x}}}{2}\,,
\ee
which follows from $S_{\tau}^{\tau} = 0$ in (\ref{energy_dens}), where we used $\tilde{x} \equiv \frac{\a^2(\tau)}{4\a}$, and 
 $\tilde{I}_x \equiv \sqrt{1+ \frac{m_i}{\tilde{x}^{2}} - \frac{2 q^2}{\tilde{x}^{3}} + \l}$.
The first derivative of the potential evaluated at the static radius $a$ is $V'(a)=0$, while for the second derivative we obtain
\be \lab{ddV_throat}
V''(a) = \frac{3 q^2 - 4 k x^2}{4 \a x^2 (k + x)} \,,
\ee
where the prime stands for the derivative with respect to the radius of the shell.
Evaluating the latter for the wormhole configurations, with parameters given as in (\ref{q_x}) and (\ref{mi_x}) in the throat, the second derivative of the potential at the equilibrium position is
\be
V''(a) = \frac{8 + 9 x}{4 \a (1 + x)} > 0 \,, \mbox{ $q \neq 0$, $\l = 0$, $k=1$,} 
\ee
yielding stable symmetric vacuum throats in every case, cf. \cite{Kokubu:2015spa}. The complete stability of shells in the wormhole we are interested in will exclusively depend on stability of the bubbles. 
\footnote{For completeness, we mention here that, including cosmological constant solutions, the general symmetric throat has
$\l =  3 + 4k/x - q^2/x^3 $, $m_i = 4 k^2+ 4 k x + 3 q^2/x$ and is stable if $3 q^2 > 4 k x^2$, as it is given by (\ref{ddV_throat}) by using the general condition $k+x>0$.
We note that if there were vanishing charge, the stability would depend exclusively on the value of the curvature $k$. 
Particularly, if $q=0$ and $k=1$ solutions exist only for $\l>0$ and are unstable under radial perturbations.}


Dynamics of the bubbles with radius $\tilde{b} \equiv b(\tau)$ is described by $\dot{\tilde{b}}^2+ V(\tilde{b}) = 0$, with the effective potential
\be
V(\tilde{b}) = k + \tilde{y}  -  \frac{\tilde{y}}{3}  \lp \tilde{I}_y - \tilde{E}_y  + \frac{ \tilde{I}_y  \tilde{E}_y}{ \tilde{I}_y - \tilde{E}_y } \rp
\ee 
where the dynamical parameter is $\tilde{y} \equiv \frac{b^2(\tau)}{4\a}$. 
Evaluating at the static position of radius $b$ we have: $\tilde{y} = y$, $\tilde{I}_{y} = {I}_{y}$ and $\tilde{E}_{y} = {E}_{y}$.
The first derivative of the potential at the equilibrium position is $V'(b)=0$, while for the second derivative we obtain
\be
V''(b) = \frac{1}{\a} 	\lp \frac{3 \, q^2/y^3}{I_{y} - E_{y}} - \frac{I_{y} - E_{y}}{I_{y} E_{y}}  - 1 \rp \,.
\ee
The stability requirement $V''(b)>0$ can be expressed generically as
\be \lab{stability_b}
9 q^2 (k+y) - (\z_y + y) \lbr 3 q^2 + y^2 (3 (k+y) - \z_y)\rbr  >0
\ee
where we used equations (\ref{static_pot_y}) and (\ref{zeda_y}), together with the general condition $\z_y + y > 0$ for vacuum static bubbles. 
From the latter, a general feature can be mentioned; bubbles are unstable without electromagnetic field for any possible value of cosmological constant. 
Then, only charged configurations would admit stable vacuum bubbles.

\begin{figure} [h!] 
\centering
\captionsetup[subfigure]{margin={0.8cm,0.5cm},justification=centering}
\subfloat[Charge against external mass.] 
{\label{Q_vs_M}\includegraphics[width=0.49\textwidth]{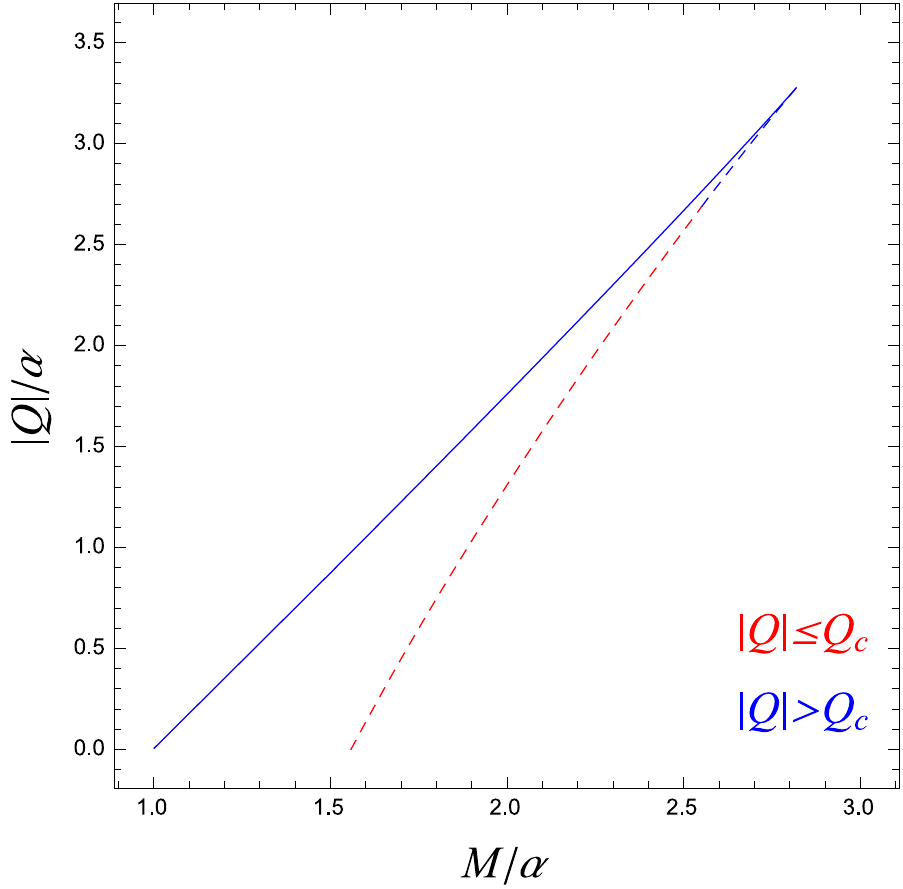}}
 \hspace{0.01cm}
\subfloat[Bubble radii (upper curve) and throat radius (lower curve).]
  {\label{ab_vs_M}
    \includegraphics[width=0.49\textwidth]{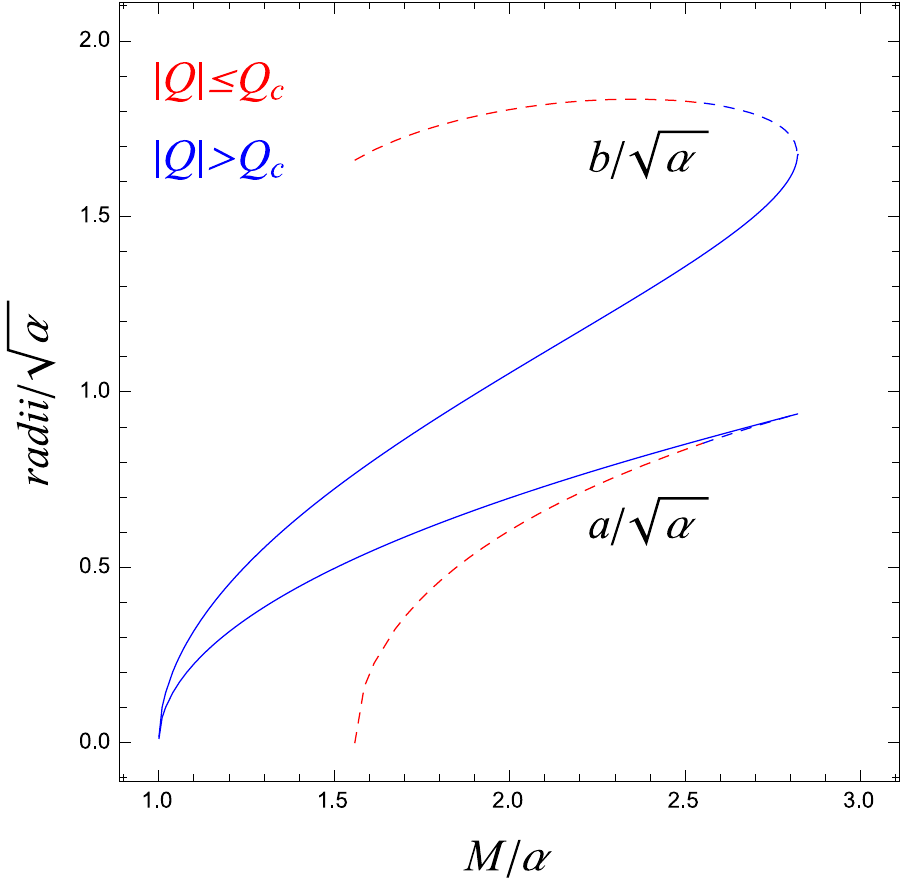}}
\caption{{Solid curves represent stable configurations (only possible with supercritical charge values). Dashed curves are unstable solutions.}}
      \label{vs_M} 
\end{figure}

The complete stability of the shells in our wormhole is therefore obtained from the latter inequality with $k=1$ and $\z_y = q^2/y^2$. 
%
The space of solutions 
for the electro-vacuum wormhole spacetime with flat asymptotics, together with the stability condition, are shown in the following figures.
Figure \ref{vs_a} shows the dimensionless bubble radius and dimensionless mass parameters, against dimensionless throat radius $a/\sqrt{\alpha}$. Figure \ref{vs_M} shows the dimensionless charge and shells radii against dimensionless external mass $M/\alpha$.
The curves in the different sets of parameter axes represent the wormhole solutions. 
Solid line curves represent stable configurations, dashed lines are unstable. 
Stable configurations occur only with charges greater than the critical value given by the external asymptotically flat black hole metric, the cases with $|Q| > Q_c$ are painted in blue, while $|Q| \leqslant Q_c$ are in red. 
 \begin{figure} [h!] 
\centering
\captionsetup[subfigure]{margin={0.8cm,0.5cm},justification=centering}
\subfloat[Bubble radius against throat radius.] 
{\label{b_vs_a}\includegraphics[width=0.481\textwidth]{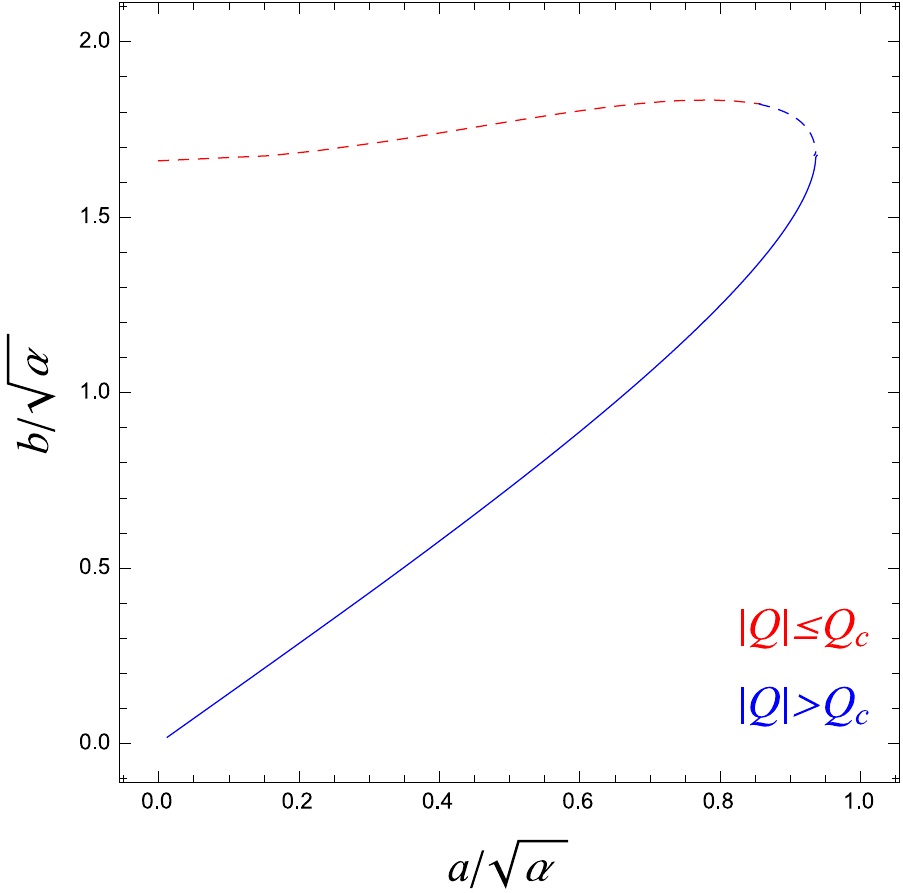}}
 \hspace{0.01cm}
\subfloat[Interior mass (upper curve) and exterior mass (lower curve).] 
  {\label{M_vs_a}
    \includegraphics[width=0.475\textwidth]{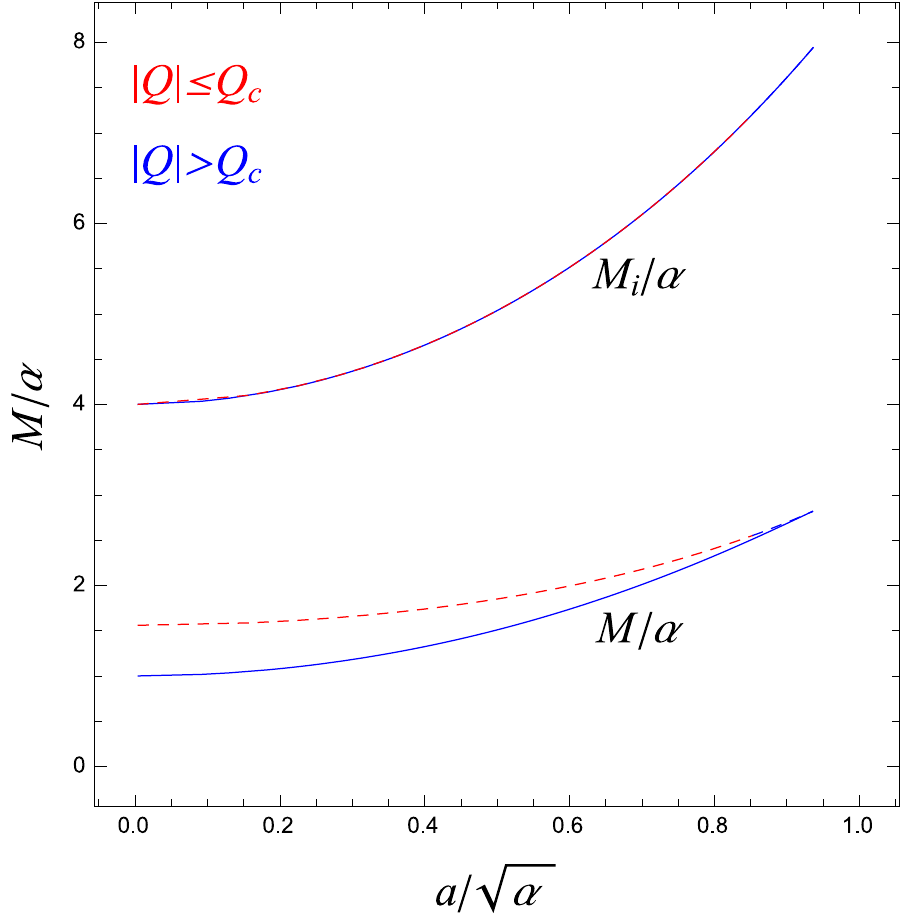}}
\caption{{Solid curves represent stable configurations (only possible with supercritical charge values). Dashed curves are unstable solutions.}}
      \label{vs_a} 
    \end{figure}
    
The neutral equilibrium condition $V''(b)=0$ for the configuration space of the bubble can be obtained by equating to zero the left hand side of (\ref{stability_b}) with $k=1$ and $\l=0$. This gives the relation between $y$ and the dimensionless charge $q$ in the configuration of neutral equilibrium of the bubble, namely
\be
q^2_{ne}= \frac{y_{ne}^2}{2} \lp 3 + 2 y_{ne} - \sqrt{9 + 6 y_{ne} - 2 y_{ne}^2} \rp \,.
\ee
By introducing this charge in the formulas for the static configuration (\ref{mi_yq}) and (\ref{m_yq}) we would obtain the mass parameters for the neutral equilibrium curves of the bubble; these are $m = m(y_{ne})$ and $m_i = m_i(y_{ne})$. 
On the other hand considering equations (\ref{q_x}) and (\ref{mi_x}) for the static throat of our wormhole we can write the charge in terms of interior mass as
\be
q^2 = \frac{1}{243} \lbr (28 + 9 m_i)^{3/2} - 80 - 108 m_i \rbr 
\ee
for $m_i >4$. Requiring these two square charges to be the same, by evaluating the latter at $m_i=m_i(y_{ne})$, we achieve the neutral equilibrium point in the wormhole space of solutions. Solving accordingly we obtain that $y_{ne}\simeq 0.70$ and, for the original metric parameters at the neutral equilibrium point, we have:
$a_{ne} \simeq 0.94 \sqrt{\alpha}$, 
$b_{ne} \simeq1.68\sqrt{\alpha}$, 
$|Q|_{ne} \simeq 3.28 \alpha $, 
$M_{ne} \simeq 2.82 \alpha$, and 
${M_i}_{ne} \simeq 7.94 \alpha $. Each of these neutral equilibrium parameters can be read in the corresponding plotted figure.
Stable wormholes occur with bubble radius $b<b_{ne}$ and charges greater than the critical charge of the exterior black hole manifolds. The corresponding mass and charge for these solutions ranges in $\alpha<M<M_{ne}$, and $0<|Q|<|Q|_{ne}$, respectively.

From the stability analysis, we conclude that the $U(1)$ charge $Q$ is what stabilizes the wormhole solution under small scalar perturbations, cf. \cite{Garraffo:2007fi}. It is worth pointing out that the stable solutions require supercritical values $|Q|>Q_c$. This might be relevant for some questions on quantum gravity: In the path integral approach, one formally defines the theory as the fluctuations about saddle points that obey certain asymptotic conditions that define charges at infinity. One picks such saddles from a set of physically sensible classical solutions, typically excluding naked singularities and other pathological configurations. This means that certain regions of the space of charges will be in principle excluded. For example, solutions (\ref{metric})-({\ref{BD}}) with $|Q|>Q_c$ are expected to develop a branch singularity and therefore one excludes such values of the charge. Here, we are finding that completely regular solutions --with two asymptotic regions -- actually exist for supercritical values of the charge, and thus such bound should be reconsidered. Conversely, the analysis above also tells us that wormhole solutions with $|Q|<Q_c$ are unstable, and this means something about for the uniqueness of the solution (\ref{metric})-({\ref{BD}}) for a given set of variables $M$ and $|Q|<Q_c$.

\section{Anti-de Sitter space}
\lab{AdS}
Wormhole solutions also exist in (A)dS spacetime provided $\Lambda\neq0$. Probably, the most interesting examples are the AdS wormholes, as they allow to think of their implications for AdS/CFT. These geometries have two asymptotically AdS$_5$ regions of effective cosmological constant  
\begin{equation}
\Lambda_{\text{eff}} = \frac{3}{2\alpha}( \sqrt{1+\lambda }-1) . 
\end{equation}

There are several questions one can ask about the interest of these solutions in the context of holography: First, one could ask whether the theory admits locally AdS$_5$ solutions with flat base manifolds (i.e. $k=0$). These solutions correspond to the boundary being locally $\mathbb{R}^{1,3}$. Interestingly, the theory (\ref{Uno}) with $\Lambda <0$ does admit\footnote{The case $k=+1$ is possible with both positive and negative $\Lambda $. The cases $k=0$ and $k=-1$ are only possible with $\Lambda <0$.} both planar ($k=0$) and hyperbolic ($k=-1$) wormhole solutions. In these cases, condition (\ref{pres_b}) at the vacuum bubbles is only compatible for negative $\Lambda$, while charge $Q$ is, again, an essential ingredient to satisfy equation (\ref{pres_a}) in a planar wormhole throat. The locally AdS solution with $k=1$ is also possible; it requires the presence of a non-zero electromagnetic field, as it follows from (\ref{pres_a}). 

A second question one can ask is whether the wormholes solutions are allowed by the so-called causality bounds. To answer this question one has to compare the range of the dimensionless parameter $\lambda$ where the stable AdS$_5$ wormholes are possible with the causality segment\footnote{It is worth comparing with reference \cite{Brigante:2007nu}, where the cosmological constant is denoted $\Lambda =-6/L^2$ and the coupling constant of the Gauss-Bonnet quadratic terms is denoted $\alpha = \lambda_{GB}L^2/2$. Here, we define the parameter $\lambda \equiv 4\Lambda\alpha /3$, i.e. $\lambda =-4\lambda_{GB}$. Therefore, the causality bounds on the Gauss-Bonnet coupling coming from AdS$_5$/CFT$_4$ reads (\ref{La43}); see for instance the bound after Eq. (4.9) of \cite{Brigante:2007nu}, which translates into the lower bound in (\ref{La43}). In the notation of \cite{Brigante:2007nu}, the Chern-Simons point $\lambda =-1$ reads $\lambda_{GB}=1/4$.}
\be
-\frac{9}{25} < \lambda < \frac{7}{9} \label{La43}
\ee 
that comes from AdS$_5$/CFT$_4$. It turns out that stable wormhole solutions can be seen to exist within the range $-1 \leqslant \lambda \lesssim-0.83$, which means that the causality bound excludes such solutions. An interesting particular case, which can not be excluded by the standard causality arguments, is $\lambda =-1$. At this point of the parameter space, the theory can be written as 5-dimensional $SO(4,2)$ Chern-Simons gauge theory, it has a unique maximally symmetric vacuum, and it also exhibits other special features. At this point, there are no local degrees of freedom around the vacuum \cite{Zanelli:2005sa}, and this is why it can not be excluded by the standard causality arguments. AdS$_5$ wormholes at $\lambda = -1$ have been previously considered in the context of AdS$_5$/CFT$_4$; for instance, in references \cite{Ali:2009ky, Ali:2013jya}. Traversable AdS$_5$ wormholes are dual to a pair of CFT$_4$'s interacting with each other. In \cite{Ali:2009ky, Ali:2013jya}, configurations in which a pair of charges is present in each copy of the CFT$_4$ were considered. A phase transition is seen to occur when the particles of each pair are separated from each other. A similar phenomenon is expected to occur in the wormhole geometries we constructed here, although there are some differences: Appropriate boundary conditions for the string configuration have to be considered in the non-differentiable junctions $\Sigma^b_{L,R}$ and $\Sigma^a$, and the effect of the non-vanishing charge has to be taken into account. We plan to investigate the holographic interpretation of the wormhole solution at $\lambda =-1$ in the future.

\subsection*{Acknowledgments}

The authors thank Mariano Chernicoff and Julio Oliva for discussions. This work has been supported by CONICET and University of Buenos Aires FCEyN-UBA. 



\begin{thebibliography}{100}

\bibitem{Misner:1957mt} 
  C.~W.~Misner and J.~A.~Wheeler,
  ``Classical physics as geometry: Gravitation, electromagnetism, unquantized charge, and mass as properties of curved empty space,''
  Annals Phys.\  {\bf 2}, 525 (1957).


\bibitem{Morris:1988cz} 
  M.~S.~Morris and K.~S.~Thorne,
  ``Wormholes in space-time and their use for interstellar travel: A tool for teaching general relativity,''
  Am.\ J.\ Phys.\  {\bf 56}, 395 (1988).


\bibitem{Fuller:1962zza} 
  R.~W.~Fuller and J.~A.~Wheeler,
  ``Causality and Multiply Connected Space-Time,''
  Phys.\ Rev.\  {\bf 128}, 919 (1962).


\bibitem{Maldacena:2013xja} 
  J.~Maldacena and L.~Susskind,
  ``Cool horizons for entangled black holes,''
  Fortsch.\ Phys.\  {\bf 61}, 781 (2013),
  [arXiv:1306.0533 [hep-th]].


\bibitem{Chernicoff:2013iga} 
  M.~Chernicoff, A.~G\"uijosa and J.~F.~Pedraza,
  ``Holographic EPR Pairs, Wormholes and Radiation,''
  JHEP {\bf 1310}, 211 (2013)
  [arXiv:1308.3695 [hep-th]].
	
	
\bibitem{Gao:2016bin} 
  P.~Gao, D.~L.~Jafferis and A.~Wall,
  ``Traversable Wormholes via a Double Trace Deformation,''
  JHEP {\bf 1712}, 151 (2017),
  [arXiv:1608.05687 [hep-th]].
	

\bibitem{Maldacena:2017axo} 
  J.~Maldacena, D.~Stanford and Z.~Yang,
  ``Diving into traversable wormholes,''
  Fortsch.\ Phys.\  {\bf 65}, no. 5, 1700034 (2017),
  [arXiv:1704.05333 [hep-th]].

\bibitem{Maldacena:2018lmt} 
  J.~Maldacena and X.~L.~Qi,
  ``Eternal traversable wormhole,''
  [arXiv:1804.00491 [hep-th]].

\bibitem{Gao:2018yzk} 
  P.~Gao and H.~Liu,
  ``Regenesis and quantum traversable wormholes,''
  [arXiv:1810.01444 [hep-th]].

\bibitem{Caceres:2018ehr} 
  E.~Caceres, A.~S.~Misobuchi and M.~L.~Xiao,
  ``Rotating traversable wormholes in AdS,''
  JHEP {\bf 1812}, 005 (2018)
  [arXiv:1807.07239 [hep-th]].

\bibitem{Betzios:2019rds} 
  P.~Betzios, E.~Kiritsis and O.~Papadoulaki,
  ``Euclidean Wormholes and Holography,''
  [arXiv:1903.05658 [hep-th]].

\bibitem{Freivogel:2019lej}
  B.~Freivogel, V.~Godet, E.~Morvan, J.~F.~Pedraza and A.~Rotundo,
  ``Lessons on Eternal Traversable Wormholes in AdS,''
  [arXiv:1903.05732 [hep-th]].
		
\bibitem{Maldacena:2018gjk}
  J.~Maldacena, A.~Milekhin and F.~Popov,
  ``Traversable wormholes in four dimensions,''
  [arXiv:1807.04726 [hep-th]].

\bibitem{Ayon-Beato:2015eca} 
  E.~Ayon-Beato, F.~Canfora and J.~Zanelli,
  ``Analytic self-gravitating Skyrmions, cosmological bounces and AdS wormholes,''
  Phys.\ Lett.\ B {\bf 752}, 201 (2016)
  [arXiv:1509.02659 [gr-qc]].

\bibitem{Anabalon:2018rzq} 
  A.~Anabal\'on and J.~Oliva,
  ``Four-dimensional Traversable Wormholes and Bouncing Cosmologies in Vacuum,''
  JHEP {\bf 1904}, 106 (2019),
  [arXiv:1811.03497 [hep-th]].

\bibitem{Anabalon:toappear} 
  A.~Anabal\'on, B. de Wit and J.~Oliva,
  ``Supersymmetric traversable wormholes in four space-time dimensions,''
  to appear.

\bibitem{Horowitz:2019hgb} 
  G.~T.~Horowitz, D.~Marolf, J.~E.~Santos and D.~Wang,
  ``Creating a Traversable Wormhole,''
 [arXiv:1904.02187 [hep-th]].

\bibitem{Graham:2007va} 
  N.~Graham and K.~D.~Olum,
  ``Achronal averaged null energy condition,''
  Phys.\ Rev.\ D {\bf 76}, 064001 (2007)
  [arXiv:0705.3193 [gr-qc]].

\bibitem{Dotti:2006cp}
  G.~Dotti, J.~Oliva and R.~Troncoso,
  ``Static wormhole solution for higher-dimensional gravity in vacuum,''
  Phys.\ Rev.\ D {\bf 75}, 024002 (2007),
  [hep-th/0607062].

\bibitem{Dotti:2007az} 
  G.~Dotti, J.~Oliva and R.~Troncoso,
  ``Exact solutions for the Einstein-Gauss-Bonnet theory in five dimensions: Black holes, wormholes and spacetime horns,''
  Phys.\ Rev.\ D {\bf 76}, 064038 (2007),
  [arXiv:0706.1830 [hep-th]].

\bibitem{Correa:2008nq} 
  D.~H.~Correa, J.~Oliva and R.~Troncoso,
  ``Stability of asymptotically AdS wormholes in vacuum against scalar field perturbations,''
  JHEP {\bf 0808}, 081 (2008)
  [arXiv:0805.1513 [hep-th]].

\bibitem{Zanelli:2005sa} 
  J.~Zanelli,
  ``Lecture notes on Chern-Simons (super-)gravities. Second edition (February 2008),''
  [hep-th/0502193].


\bibitem{Maeda:2008nz} 
  H.~Maeda and M.~Nozawa,
  ``Static and symmetric wormholes respecting energy conditions in Einstein-Gauss-Bonnet gravity,''
  Phys.\ Rev.\ D {\bf 78}, 024005 (2008),
  [arXiv:0803.1704 [gr-qc]].
	
\bibitem{Kokubu:2015spa} 
  T.~Kokubu, H.~Maeda and T.~Harada,
  ``Does the Gauss-Bonnet term stabilize wormholes?,''
  Class.\ Quant.\ Grav.\  {\bf 32}, no. 23, 235021 (2015)
  [arXiv:1506.08550 [gr-qc]].
	
\bibitem{Brigante:2007nu}
  M.~Brigante, H.~Liu, R.~C.~Myers, S.~Shenker and S.~Yaida,
  ``Viscosity Bound Violation in Higher Derivative Gravity,''
  Phys.\ Rev.\ D {\bf 77}, 126006 (2008),
  [arXiv:0712.0805 [hep-th]].

\bibitem{Hofman:2009ug} 
  D.~M.~Hofman,
  ``Higher Derivative Gravity, Causality and Positivity of Energy in a UV complete QFT,''
  Nucl.\ Phys.\ B {\bf 823}, 174 (2009),
  [arXiv:0907.1625 [hep-th]].


\bibitem{Antoniadis:1997eg} 
  I.~Antoniadis, S.~Ferrara, R.~Minasian and K.~S.~Narain,
  ``R**4 couplings in M and type II theories on Calabi-Yau spaces,''
  Nucl.\ Phys.\ B {\bf 507}, 571 (1997)
  [hep-th/9707013].



\bibitem{Ferrara:1996hh} 
  S.~Ferrara, R.~R.~Khuri and R.~Minasian,
  ``M theory on a Calabi-Yau manifold,''
  Phys.\ Lett.\ B {\bf 375}, 81 (1996)
  [hep-th/9602102].

\bibitem{Guica:2005ig} 
  M.~Guica, L.~Huang, W.~Li and A.~Strominger,
  ``R**2 corrections for 5-D black holes and rings,''
  JHEP {\bf 0610}, 036 (2006)
  [hep-th/0505188].


\bibitem{Lanczos:1938sf} 
  C.~Lanczos,
  ``A Remarkable property of the Riemann-Christoffel tensor in four dimensions,''
  Annals Math.\  {\bf 39}, 842 (1938).

\bibitem{Lovelock:1972vz} 
  D.~Lovelock,
  ``The four-dimensionality of space and the Einstein tensor,''
  J.\ Math.\ Phys.\  {\bf 13}, 874 (1972).

\bibitem{Zwiebach:1985uq} 
  B.~Zwiebach,
  ``Curvature Squared Terms and String Theories,''
  Phys.\ Lett.\  {\bf 156B}, 315 (1985).

\bibitem{Camanho:2014apa} 
  X.~O.~Camanho, J.~D.~Edelstein, J.~Maldacena and A.~Zhiboedov,
  ``Causality Constraints on Corrections to the Graviton Three-Point Coupling,''
  JHEP {\bf 1602}, 020 (2016),
  [arXiv:1407.5597 [hep-th]].


\bibitem{Camanho:2012da} 
  X.~O.~Camanho, J.~D.~Edelstein, G.~Giribet and A.~Gomberoff,
  ``A New type of phase transition in gravitational theories,''
  Phys.\ Rev.\ D {\bf 86}, 124048 (2012)
  [arXiv:1204.6737 [hep-th]].


\bibitem{Izumi:2014loa} 
  K.~Izumi,
  ``Causal Structures in Gauss-Bonnet gravity,''
  Phys.\ Rev.\ D {\bf 90}, no. 4, 044037 (2014)
  [arXiv:1406.0677 [gr-qc]].


\bibitem{Reall:2014pwa} 
  H.~Reall, N.~Tanahashi and B.~Way,
  ``Causality and Hyperbolicity of Lovelock Theories,''
  Class.\ Quant.\ Grav.\  {\bf 31}, 205005 (2014)
  [arXiv:1406.3379 [hep-th]].



\bibitem{Lovelock:1971yv}
  D.~Lovelock,
  ``The Einstein tensor and its generalizations,''
  J.\ Math.\ Phys.\  {\bf 12}, 498 (1971).





\bibitem{Garraffo:2007fi}
  C.~Garraffo, G.~Giribet, E.~Gravanis and S.~Willison,
  ``Gravitational solitons and C0 vacuum metrics in five-dimensional Lovelock gravity,''
  J.\ Math.\ Phys.\  {\bf 49}, 042502 (2008),
  [arXiv:0711.2992 [gr-qc]].

\bibitem{Richarte:2007zz}
  M.~G.~Richarte and C.~Simeone,
  ``Thin-shell wormholes supported by ordinary matter in Einstein-Gauss-Bonnet gravity,''
  Phys.\ Rev.\ D {\bf 76}, 087502 (2007),
  Erratum: Phys.\ Rev.\ D {\bf 77}, 089903 (2008),
  [arXiv:0710.2041 [gr-qc]].

\bibitem{Simeone:2011zz} 
  C.~Simeone,
  ``Addendum to `Thin-shell wormholes supported by ordinary matter in Einstein-Gauss-Bonnet gravity',''
  Phys.\ Rev.\ D {\bf 83}, 087503 (2011),
  [arXiv:1203.3755 [gr-qc]].


\bibitem{Belin:2019mnx}
  A.~Belin, D.~M.~Hofman and G.~Mathys,
  ``Einstein gravity from ANEC correlators,''
  [arXiv:1904.05892 [hep-th]].


\bibitem{Boulware:1985wk} 
  D.~G.~Boulware and S.~Deser,
  ``String Generated Gravity Models,''
  Phys.\ Rev.\ Lett.\  {\bf 55}, 2656 (1985).
	
\bibitem{Wiltshire:1988uq} 
  D.~L.~Wiltshire,
  ``Black Holes in String Generated Gravity Models,''
  Phys.\ Rev.\ D {\bf 38}, 2445 (1988).



\bibitem{Poisson:1995sv} 
  E.~Poisson and M.~Visser,
  ``Thin shell wormholes: Linearization stability,''
  Phys.\ Rev.\ D {\bf 52}, 7318 (1995),
  [gr-qc/9506083].

\bibitem{Davis:2002gn} 
  S.~C.~Davis,
  ``Generalized Israel junction conditions for a Gauss-Bonnet brane world,''
  Phys.\ Rev.\ D {\bf 67}, 024030 (2003),
  [hep-th/0208205].


\bibitem{Garraffo:2008hu} 
  C.~Garraffo and G.~Giribet,
  ``The Lovelock Black Holes,''
  Mod.\ Phys.\ Lett.\ A {\bf 23}, 1801 (2008)
  [arXiv:0805.3575 [gr-qc]].



\bibitem{Visser:2003ge} 
  M.~Visser and D.~L.~Wiltshire,
  ``Stable gravastars: An Alternative to black holes?,''
  Class.\ Quant.\ Grav.\  {\bf 21}, 1135 (2004)
  [gr-qc/0310107].


\bibitem{Brady:1991np} 
  P.~R.~Brady, J.~Louko and E.~Poisson,
  ``Stability of a shell around a black hole,''
  Phys.\ Rev.\ D {\bf 44}, 1891 (1991).



\bibitem{Zanelli:2005sa} 
  J.~Zanelli,
  ``Lecture notes on Chern-Simons (super-)gravities. Second edition (February 2008),''
  [hep-th/0502193].


\bibitem{Lu:2013awa} 
  H.~L\"u, J.~F.~Vazquez-Poritz and Z.~Zhang,
  ``Strings on AdS Wormholes and Nonsingular Black Holes,''
  Class.\ Quant.\ Grav.\  {\bf 32}, no. 2, 025005 (2015),
  [arXiv:1309.2957 [hep-th]].

\bibitem{Ali:2009ky} 
  M.~Ali, F.~Ruiz, C.~Saint-Victor and J.~F.~Vazquez-Poritz,
  ``Strings on AdS Wormholes,''
  Phys.\ Rev.\ D {\bf 80}, 046002 (2009)
  [arXiv:0905.4766 [hep-th]].

\bibitem{Ali:2013jya} 
  M.~Ali, F.~Ruiz, C.~Saint-Victor and J.~F.~Vazquez-Poritz,
  ``Strings on AdS wormholes,''
  J.\ Phys.\ Conf.\ Ser.\  {\bf 462}, no. 1, 012058 (2013).
	
	






\end{thebibliography}
\end{document}